\documentclass[twocolumn,twoside,slac_two]{revtex4}
\usepackage{graphicx}
\usepackage{fancyhdr}
\usepackage{bm}
\begin{document}

\title{\boldmath Spectroscopy
of Hadrons
with $b$ Quarks \unboldmath} 


\author{R. Van Kooten}
\affiliation{Indiana University, Bloomington, IN, U.S.A.}

\begin{abstract}
Recent experimental results on the spectroscopy of $B$ hadrons are presented.
The focus is primarily on the heavier states currently accessible only
at the Tevatron and the $B$ factories running at the 
$\Upsilon(5S)$ resonance, i.e., $B_s^*$, the orbitally excited states
$B^{**}$ and $B^{**}_s$, and the new $b$-baryon states $\Sigma^{(*)}_b$,
$\Xi_b$, and $\Omega_b$.
\end{abstract}

\maketitle

\thispagestyle{fancy}


\section{Introduction}
Heavy quark hadrons are the hydrogen atom of QCD, and 
$b$ hadrons offer the heaviest quarks that form bound systems.
The study of their spectroscopy provides  sensitive tests of
potential models, Heavy Quark Effective Theory (HQET~\cite{HQET}), 
and all regimes of QCD in general, including
the non-perturbative regime described by lattice gauge 
calculations~\cite{lattice}.

There has been somewhat of a ``renaissance" of new heavy flavor
spectroscopy results these past few years with new data coming from 
the $B$ factories and the Tevatron.  This contribution focuses
on new experimental results from the Tevatron and the
$B$ factories running on the $\Upsilon(5S)$ resonance, although a very 
recent result on the $B^0$-$B^+$ mass difference from the BaBar
Collaboration is included for completeness.

The Tevatron has the capability of producing heavier states not
accessible at the $B$ factories running at the $\Upsilon(4S)$, i.e., 
the heavy $B$  mesons: $B^0_s$ ($\bar{b}s$, the ground state  
with the spins of the quarks anti-aligned), 
$B^*_s$ ($\bar{b} s$, with the spins of the quarks
aligned), $B_c$ ($\bar{b} c$, the ground state), 
$B^{**}$ ($\bar{b} d$, with the quarks having relative orbital
angular momentum), and $B_s^{**}$ ($\bar{b}s$, with the quarks having
relative orbital angular momentum);
and the heavy $b$ baryons: $\Lambda_b^0$ ($bud$), 
$\Sigma_b^{(*)\pm}$ ($buu$ and $bdd$), 
and $\Xi_b$ ($bsd$), with many more baryonic states possible from
the remaining combinations of quarks.
Properties of the $B^0_s$ are covered by other
contributions to this conference~\cite{fpcp08}.
These results are complementary to those from the $\Upsilon(4S)$
$B$ factories, but there are exciting new results on $B_s^*$  from
some luminosity collected on the $\Upsilon(5S)$ resonance
by the CLEO and Belle Collaborations.

Charge conjugate modes and reactions are always implied in this
contribution.

\section{\boldmath $B^0$-$B^+$ Mass Difference}

The $B^0$-$B^+$ mass difference probes the size of Coulomb 
contributions to the quark structure of mesons. Precise predictions
of this mass split are fairly uncertain since contributions from the
quark-mass difference $m(d) - m(u)$ and from Coulomb effects have
similar magnitudes and opposite signs. 

\begin{figure}[h]
\centering
\includegraphics[width=80mm]{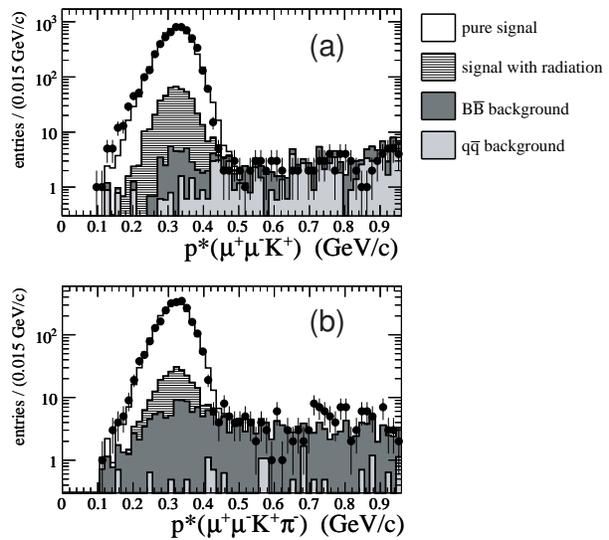}
\caption{(a) Reconstructed momenta of
$B^+ \rightarrow J/\psi K^+$
candidates
and (b) reconstructed momenta
of the similar topology 
$B^0 \rightarrow J/\psi K^+ \pi^-$ by the BaBar Collaboration.}
\label{fig:babar}
\end{figure}

The BaBar Collaboration has made a precision measurement of this
mass splitting~\cite{babar_mass_split} by making  full 
reconstructions of the two $B$ mesons with similar final states:
$B^+ \rightarrow J/\psi K^+ \rightarrow \mu^+ \mu^-$ and
$B^0 \rightarrow J/\psi K^+ \pi^- \rightarrow \mu^+ \mu^- K^+ \pi^-$.
By carefully comparing the two reconstructed momenta distributions
as shown in Fig.~\ref{fig:babar}, they have measured
\begin{equation}
\Delta M_B = M(B^0) - M(B^+) = +0.33 \pm 0.05 \pm 0.03 
\thinspace {\mathrm{MeV}},
\nonumber
\end{equation}
which is consistent with, but much more precise than the PDG 2008
world average of 
$+0.37 \pm 0.24 \thinspace {\mathrm{MeV}}$~\cite{PDG08}.

\section{\boldmath Mass of $B_c$ Meson \unboldmath}

One of the more interesting mesons that can be studied at the Tevatron
is the $B_c$ ($\bar{b}c$) meson. 
It is unique in that it contains two {\em different} heavy quarks,
both with relatively large widths to decay.
While charmonium ($c\bar{c}$) or bottomonium ($b\bar{b}$) mesons 
are also interesting with two heavy quarks,
these both decay strongly, whereas the $B_c $ meson decays 
weakly. The $B_c$ meson is expected to have the shortest 
lifetime of all
weakly decaying $b$ hadrons, 
with a predicted lifetime of about one-third
of the other $B$~mesons.
It can decay via the $b$ quark: 
$B_c \rightarrow B^0_s \pi^+, \thinspace B^0_s \ell^+ \nu$;
via the $c$ quark: 
$B_c \rightarrow J/\psi \pi^+, \thinspace
J/\psi D^+_s, J/\psi \ell^+ \nu$; or by annihilation:
$B_c \ell^+ \nu$.

First evidence for the $B_c$ meson was via its decay
into $J/\psi \ell^+ \nu$~\cite{bc_CDF_Run1}, but with the missing 
neutrino, this did not provide a very precise mass measurement.
With the greater statistics collected by the Tevatron, 
the reconstruction of the exclusive
decay $B_c \rightarrow J/\psi \pi^+$ provides for a much 
more precise mass measurement and in both cases, the
$J/\psi \rightarrow \mu^+ \mu^-$ allows for very efficient dimuon
triggering.
In similar analyses,  CDF~\cite{bc_CDF_Run2} and
D\O~\cite{bc_D0_Run2}  optimized their selection criteria
on large control samples of $B^+ \rightarrow J/\psi K^+$, 
$J/\psi \rightarrow \mu^+ \mu^-$ with a topology similar to that
of the signal channel. 

\begin{figure}[h]
\centering
\includegraphics[width=80mm]{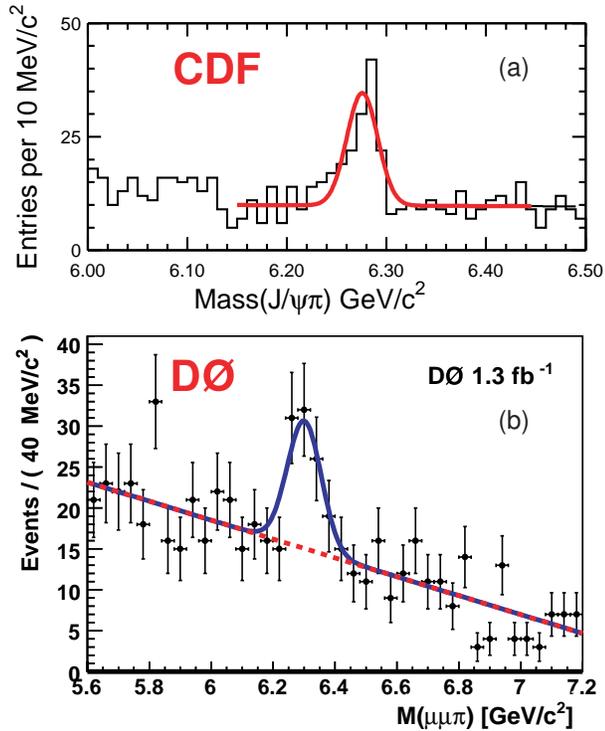}
\caption{$B_c$ candidate signals in reconstructed
$J/\psi \pi$ invariant mass distributions 
after all cuts from the described
(a) CDF and (b) D\O\ analyses.
}
\label{fig:bcmeas}
\end{figure}

\begin{figure}[h]
\centering
\includegraphics[width=80mm]{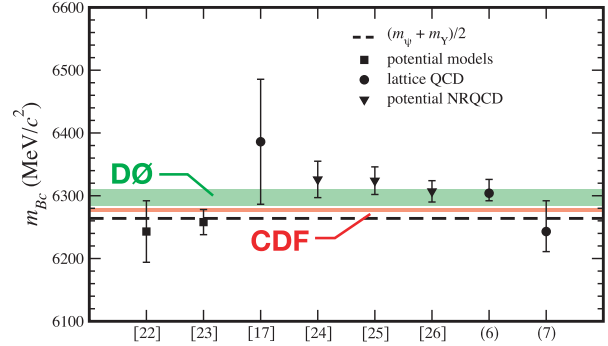}
\caption{CDF and D\O\ $B_c$ mass measurements (horizontal bands
with width equal to $\pm 1\sigma$) compared to various 
theoretical predictions as collated in Ref.~\cite{bcpreds}
(figure adapted from same reference).
}
\label{fig:bcpred}
\end{figure}

Making unbinned likelihood fits to the $J/\psi \pi$ invariant mass
distributions as shown in Fig.~\ref{fig:bcmeas},
CDF, using 2.4~fb$^{-1}$ of data finds $108 \pm 15$ 
signal candidates with a significance above background of 
greater than 8$\sigma$, and D\O, using 1.3~fb$^{-1}$ of data finds
$54 \pm 12$ signal candidates with a significance above background
of greater than 5$\sigma$. Mass measurements are made with results of:
\begin{eqnarray}
M(B_c^+) & = & 6275.6 \pm 2.9 \pm 2.5 \thinspace {\mathrm{MeV}} 
\hspace{1.5mm} {\mathrm{CDF}}, \nonumber \\
M(B_c^+) & = & 6300 \pm 14 \pm 5 \thinspace {\mathrm{MeV}}
\hspace{1.5mm} {\mathrm{D\O}}.
\end{eqnarray}
These measurements can be compared to a number of
theoretical predictions~\cite{bcpreds} as shown in 
Fig.~\ref{fig:bcpred}. These experimental measurements now
have smaller uncertainty than the corresponding uncertainties on
the theoretical predictions, indicating a need for more
theoretical work in this area.

\section{\boldmath 
Spectroscopy of Excited $B$ Mesons}

In the heavy-quark limit, the spin of the heavy quark 
decouples from the light degrees of freedom and the heavy and
light systems can be considered separately.
Furthermore, as the system has been shown to not depend,
to first order, on the flavor of the heavy quark, the
system can be defined by the quantum numbers of the 
``brown muck" of the light degrees of freedom describing
the gluons and light quarks.

Given the approximation within HQET that the heavy quark is 
at rest in the frame of the hadron, we can describe the heavy
quark through the assignment of a spin quantum number, 
$\vec{s}_Q$.  The light degrees of freedom which, following
the simple hydrogen atom model of HQET, can be thought of 
as orbiting the heavy quark are assigned a total angular
momentum $\vec{j}_q = \vec{s}_q + \vec{L}$, where $s_q$ and $L$
are the spin and orbital angular momentum of the light degrees of
freedom.  Finally the total angular momentum
is given by $\vec{J} = \vec{s}_Q + \vec{j}_q$.
$B^{**}$ and
$B^{**}_s$ mesons (also denoted $B_{J}$ and $B_{sJ}$, respectively)
are composed of
a heavy $b$  quark and a lighter down or strange quark in a
$L=1$ state of orbital momentum, with four possible states in
each case as shown in Table~\ref{tab:spect}.

\begin{table}[h]
\begin{center}
\caption{Quantum numbers of orbitally excited $B$ and $B_s$  states.}
\begin{tabular}{c|c|c}
\hline
 & \multicolumn{2}{c}{$L = 0$} \\ \hline \hline
 $j_q = \frac{1}{2}$  & $J = 0$ & $J = 1$ \\
           wide         & $B^*_0$,  $B^*_{s0}$ & $B'_1$, $B'_{s1}$ \\ \hline
 $j_q = \frac{3}{2}$  & $J = 1$ & $J = 2$ \\
          narrow        & $B_1$,  $B_{s1}$ & $B^*_2$, $B^*_{s2}$ \\ \hline
    \end{tabular}
    \label{tab:spect}
    \end{center}
    \end{table}

\begin{figure}[h]
\centering
\includegraphics[width=80mm]{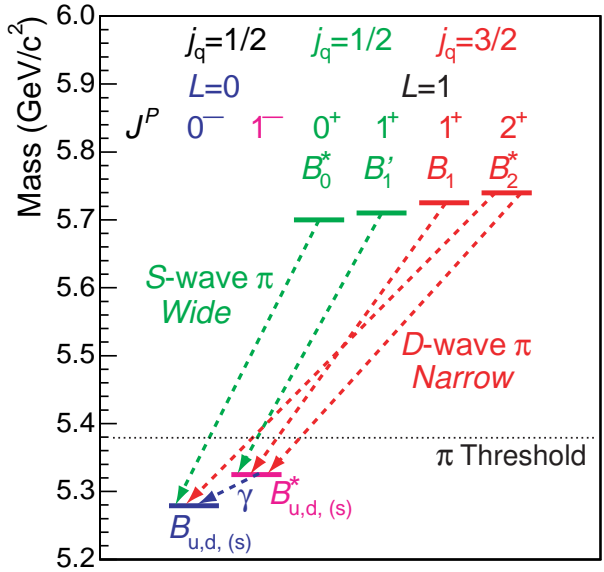}
\caption{Expected spectroscopy of the excited $B$ meson states.}
\label{fig:spect_decays}
\end{figure}

All the orbitally excited states are expected to have masses 
more than a pion mass above the ground or first excited state and 
therefore decay strongly as shown in Fig.~\ref{fig:spect_decays}. 
By conservation of parity and angular momentum,
the decay
$B^0_{(s)1} \rightarrow B^+_{(s)} \pi^-$ is not allowed.
The decay of the two  $j_q = \frac{1}{2}$ states
in each case proceeds via an $S$-wave and hence have widths
of a few hundred MeV, difficult to distinguish from
combinatorial background in effective mass spectra,
while the $j_q = \frac{3}{2}$ states,
$B_{(s)1}$ and $B^*_{(s)2}$, undergo $D$-wave decays, and due to the angular
momentum barrier have narrow widths making them experimentally 
accessible.

\subsection{\boldmath Mass of $B^*_s$ Meson \unboldmath}
      
Focusing in particular on the $B^0_s$ system, for completeness,
measurements are described of the  mass splitting between
the $B^*_s$ where the spins of the $\bar{b}$ and $s$ quarks are
aligned and the ground state $B^0_s$ with anti-aligned quark
spins. 

The $B$ factories can produce the $B^*_s$ by running
on the $\Upsilon(5S)$ resonance, i.e., 
$\Upsilon(5S) \rightarrow B^*_s \bar{B}^*_s$.

Using the kinematic variables of 
the energy difference $\Delta E \equiv E_{\mathrm{beam}} - E_B$
and the beam-constrained mass 
$M_{bc} \equiv \sqrt{E^2_{\mathrm{beam}} - \vec{p}^2_B}$, 
where $E_B$ ($\vec{p}_B$) is the energy (momentum)
of the reconstructed $B$ (or $B_s$) and 
$E_{\mathrm{beam}}$ is the beam energy, both the CLEO and
Belle Collaborations have isolated $B^*_s$ states.

\begin{figure}[h]
\centering
\includegraphics[width=80mm]{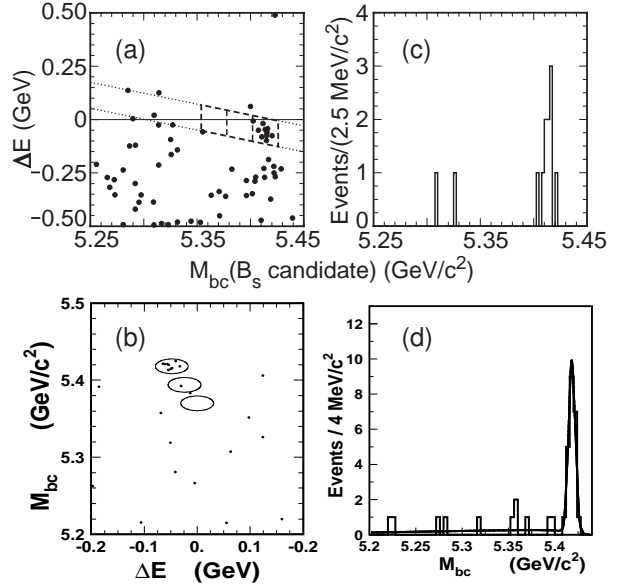}
\caption{$B^*_s$: $M_{bc}$ versus $\Delta E$ scatter plots for the
$B^0_s \rightarrow D^-_s \pi^+$ decay mode for
the (a) CLEO Collaboration and (b) Belle Collaboration.
Beam-constrained mass in the $\Delta E$ region 
where the $B^0_s$ signal from the 
$B^*_s \bar{B}^*_s$ channel is expected for (c) CLEO and
(d) Belle.}
\label{fig:bsstar}
\end{figure}

The CLEO Collaboration finds 14 candidates consistent
with $B_s$ decays into final states with a $J/\psi$ or
a $D^{(*)-}_s$~\cite{CLEO_Bs} as shown in Fig.~\ref{fig:bsstar}(a).
From an update~\cite{CLEO_B} of $M(B^*)$, they find the 
beam-constrained mass difference
$M_{bc}(B^*_s \bar{B}^*_s) - M_{bc}(B^* \bar{B}^*)$ to 
translate into the mass different 
$M(B^*_s) - M(B^*)$ and the most precise measurement of
the $B^*_s$ mass and  mass splitting: 
\begin{eqnarray}
M(B^*_s) & = &5411.7 \pm 1.6 \pm 0.6 \thinspace {\mathrm{MeV}},
\nonumber \\
\Delta M (B^*_s - B_s) & = & 45.7 \pm 1.7 \pm 0.7
\thinspace {\mathrm{MeV}}.
\end{eqnarray}

The Belle Collaboration using similar techniques~\cite{Belle_bstst}, as shown 
in Fig.~\ref{fig:bsstar}(b), find a mass
of:
\begin{equation}
M(B^*_s) = 5418 \pm 1 \pm 3 \thinspace {\mathrm{MeV}},
\end{equation}
with a 1.8$\sigma$ difference between the CLEO and Belle measurement.
The author's average of these two measurements, 
as well as the  mass splitting,
including
an older CUSB2 measurement~\cite{CUSB2}, is:
\begin{eqnarray}
M(B^*_s) & = & 5412.8 \pm 0.9  \thinspace {\mathrm{MeV}},
\nonumber \\
\Delta M (B^*_s - B_s) &  = & 46.7 \pm 1.0
\thinspace {\mathrm{MeV}}.
\end{eqnarray}

\subsection{\boldmath
Masses of $B^{**}$ Narrow States 
\unboldmath}

\begin{figure}[htb]
\centering
\includegraphics[width=80mm]{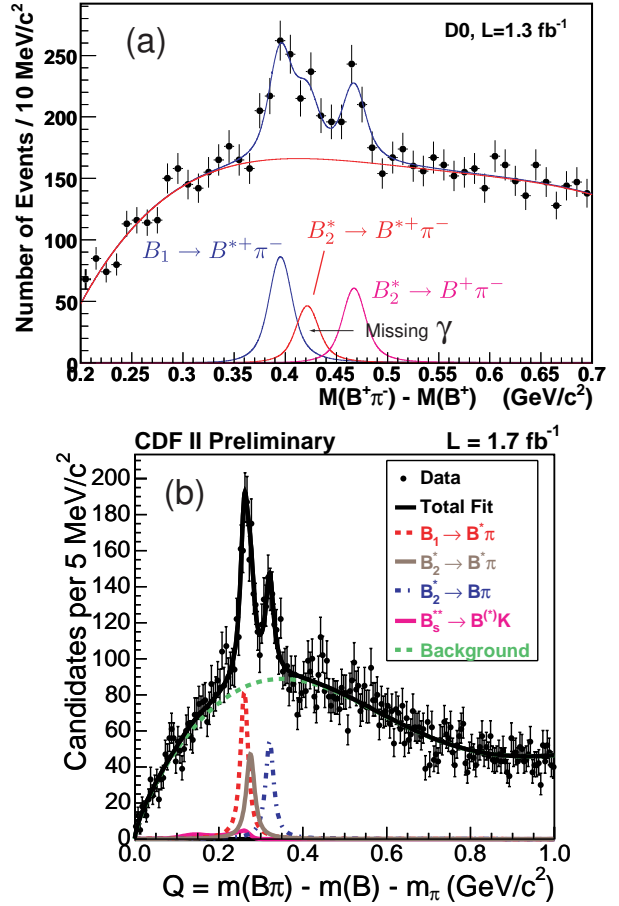}
\caption{$B^{**}$:
(a) Invariant mass difference for exclusive $B$ decays for
the D\O\ analysis with the contribution of background and
the three signal peaks shown separately; 
(b) $Q$ value for the CDF analysis along with fit contributions.
}
\label{fig:bstst}
\end{figure}

As described earlier, of the four possible $B^{**}$ states,
the $B_1$ and $B^*_2$ are narrow enough to appear as distinct
invariant mass peaks.  
Up to recently, almost all observations of the narrow states have been
made only indirectly (at LEP~\cite{PDG08}) in inclusive or semi-inclusive
decays that prevents their 
model-independent
separation and precise measurement of
properties.
Both the D\O\ and CDF Collaborations 
have reconstructed these states by first reconstructing
a $B^+ \rightarrow J/\psi K^+$ where the $J/\psi$ decays to
$\mu^+ \mu^-$ providing a solid dimuon trigger.
In the D\O\ analysis~\cite{D0_bstst}, 23k such decays
in 1.3~fb$^{-1}$ of data are 
reconstructed and then a pion is added to this combination
to search for the decay chains
$B_1 \rightarrow B^{*+} \pi^-$ and
$B^*_2 \rightarrow B^{*+} \pi^-$.
In addition, there is also the direct decay
into the ground state: $B^*_2 \rightarrow B^+ \pi^-$;
the direct decay $B_1 \rightarrow B^+ \pi^-$ is forbidden
by conservation of parity and angular momentum.
The photon from the subsequent decay $B^{*+} \rightarrow B^+ \gamma$
is {\em} not reconstructed, and as a result, invariant mass
peaks in the $M(B^+ \pi^-) - M(B^+)$ mass difference spectrum
are shifted down by 46~MeV, and three peaks are expected as
shown in Fig.~\ref{fig:bstst}(a), with two arising from the
$B^*_2$ state.
For D\O, the experimental resolution is larger than the expected line
widths of the states, and they are set to 10~MeV, but varied over 
large ranges in the systematic error determination.
This is the first time that these two states have been cleanly 
separated in a model-independent way, and the statistical significance
of the three peaks is greater than 7$\sigma$. The masses are
measured to be:
\begin{eqnarray}
M(B^0_1) & = & 5720.6 \pm 2.4 \pm 1.4 \thinspace {\mathrm{MeV}},
\nonumber \\
M(B^{*0}_2) & = & 5746.6 \pm 2.4 \pm 1.4 \thinspace {\mathrm{MeV}}.
\end{eqnarray}

The CDF Collaboration has a preliminary 
result~\cite{CDF_bstst} from a
similar analysis.  In 1.7~fb$^{-1}$ of data,  they also first
reconstruct 52k $B^+ \rightarrow J/\psi K^+$, but add
40k $B^+ \rightarrow D \pi^+$
and 11k $B^+ \rightarrow D \pi^+ \pi^- \pi^+$
candidates.
They add a pion to the reconstructed $B^+$ candidates
and observe clear peaking structure in each of
the above three channels.  The distribution of the 
$Q$ value as shown 
in Fig.~\ref{fig:bstst}(b) is fit to a similar three-peak
structure as above, but also includes peaking reflection backgrounds
from $B^{(*)}K$ candidates. Measured masses of:
\begin{eqnarray}
M(B^0_1) & = & 
5725.3^{+1.6 +1.4}_{-2.2 -1.5} \thinspace {\mathrm{MeV}}, 
\nonumber \\
M(B^{*0}_2) & = & 5740.2^{+1.7 +0.9}_{-1.8 -0.8} \thinspace {\mathrm{MeV}},
\end{eqnarray}
are the most precise to date.
The good mass resolution of the CDF detector
allows for the first preliminary measurement of the $B^{*0}_2$ width:
\begin{equation}
\Gamma(B^{*0}_2) = 22.7 ^{+3.8 +3.2}_{-3.2 -10.2}
\thinspace {\mathrm{MeV}}.
\end{equation}

\subsection{\boldmath
Masses of $B^{**}_s$ Narrow States
\unboldmath}

CDF and D\O\ have also recently measured the masses of
the narrow $B^{**}_s$ states.
Again, beginning with $B^+ \rightarrow J/\psi K^+$ 
(and CDF adds $B^+ \rightarrow D \pi^+$), 
instead of adding a pion, a kaon is added.
The decay chains of interest are therefore
$B_{s1} \rightarrow B^{*+} K^-$,
$B^*_{s2} \rightarrow B^{*+} K^-$,
and the direct decay
$B^*_{s2} \rightarrow B^+ K^-$.

\begin{figure}[htb]
\centering
\includegraphics[width=80mm]{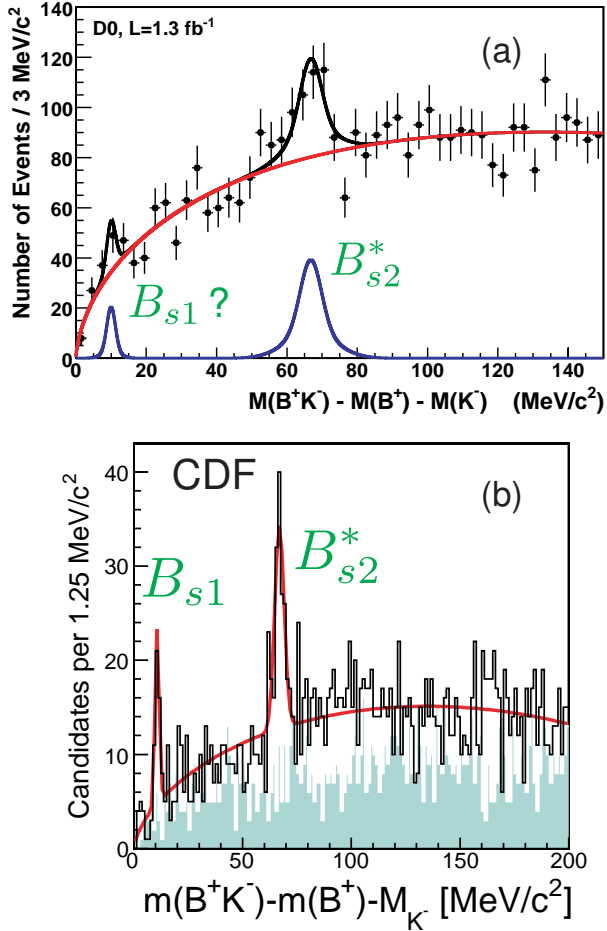}
\caption{$B^{**}_s$:
(a) Invariant mass difference for exclusive $B$ decays for
the D\O\ analysis and the fit for the two-peak hypothesis
(existence of $B_{s1}$ peak inconclusive); and
(b) $Q$ value for the CDF analysis  with fit. The filled
area is for the wrong-sign combination $B^+K^+$.
}
\label{fig:bsstst}
\end{figure}

In the analysis from the 
D\O\ Collaboration~\cite{D0_bsstst}, Fig.~\ref{fig:bsstst}(a) shows
the fit for the total number of $B^*_{s2}$ in the direct
decay $B^*_{s2} \rightarrow B^+ K^-$ with a statistical
significance greater than 5$\sigma$ in 1.3~fb$^{-1}$ with
a mass measurement of
\begin{equation}
M(B^{*0}_{s2}) = 5839.6 \pm 1.1 \pm 0.7 \thinspace
{\mathrm{MeV}}.
\end{equation}
There should be a second peak due
to decay $B^*_{s2} \rightarrow B^{*+} K^-$, shifted
down by 46~MeV in mass since the photon from the
$B^{*+}$ decay is not reconstructed.
Since the mass difference in the decay 
$B^*_{s2} \rightarrow B^{*+} K^-$ is very small,
there is a heavy phase-space suppression factor 
estimated to be 0.074; therefore, no
detectable signal is expected in the $\Delta M$ 
distribution with the current statistics.
To test for the existence of a $B_{s1}$
signal in the data, a two-peak hypothesis
is used to fit the data as shown in Fig.~\ref{fig:bsstst}(a).
The statistical significance of the $B_{s1}$ peak is less
than 3$\sigma$ so that the presence of a $B_{s1}$ state in 
the data can be neither confirmed nor excluded.

In CDF's analysis~\cite{CDF_bsstst} 
in 1~fb$^{-1}$ of data, they combine $B+$ channels
from $B^+ \rightarrow J/\psi K^+$ and 
$B^+ \rightarrow D \pi^+$.  A two-peak structure is
observed in both, and the combined fit with
excellent mass resolution is shown
in Fig.~\ref{fig:bsstst}(b). This is the first observation
of the narrow $B_{s1}$ state
with yields of $N(B_{s1}) = 36 \pm 9$ events and
$N(B*_{s2}) = 95 \pm 23$. The resulting mass measurements are:
\begin{eqnarray}
M(B^0_{s1}) & = & 5829.4 \pm 0.7 \thinspace {\mathrm{MeV}},
\nonumber \\
M(B^{*0}_{s2}) & = & 5839.6 \pm 0.7 \thinspace {\mathrm{MeV}},
\nonumber \\
\Delta M(B^{*0}_{s2} - B^0_{s1}) &  = & 10.5 \pm 0.6 
\thinspace {\mathrm{MeV}}.
\end{eqnarray}

\subsection{\boldmath
$B^{**}_{(s)}$ Summary \unboldmath}

\begin{figure}[htb]
\centering
\includegraphics[width=80mm]{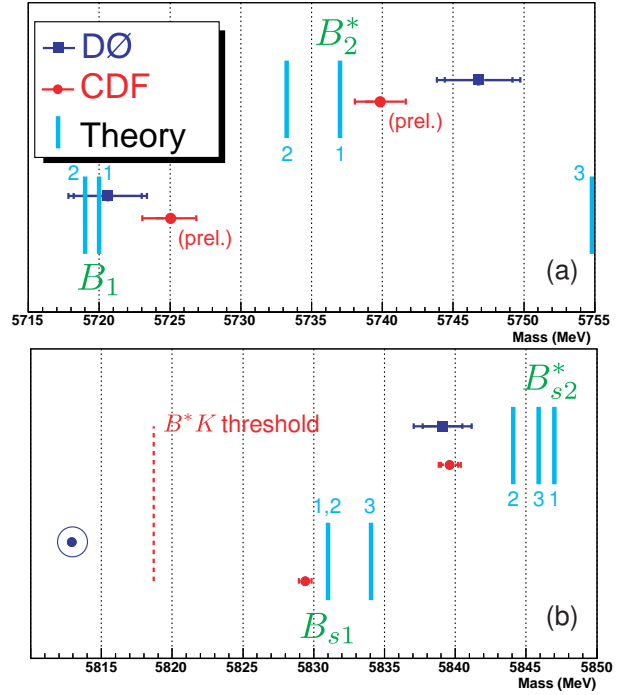}
\caption{
Summary of  mass measurements  from previous subsections
for the narrow states
(a) $B^0_1$ and $B^{*0}_2$ and
(b) $B^0_{s1}$ and $B^{*0}_{s2}$
and some representative theoretical predictions 
(1: Ref.~\cite{theory1}, 2: Ref.~\cite{theory2}, 3: Ref.~\cite{theory3}).
}
\label{fig:bss_summ}
\end{figure}

The experimental  results of the masses of
the narrow $B_{(s)J}$ states
of the previous two subsections are summarized in 
Fig.~\ref{fig:bss_summ} and compared to 
some representative theoretical predictions~\cite{theory1,theory2,theory3}.

As can be seen, there are varying degrees of agreement between
experimental measurements and predictions.  In the case of
the $B_{s1}$ state, not conclusively observed by D\O, 
if the same mass split observed in the $B^*_{2}$-$B_1$ system
is applied to the $B^{**}_s$ system, the $B_{s1}$ would in fact be
expected to be below the $B^* K$ threshold as
shown by the circled point in Fig.~\ref{fig:bsstst}(b).
More experimental
data is needed to clarify the spectra of both
the orbitally excited states of $B$ and $B_s$.

\section{\boldmath
New $b$-Flavored Baryons \unboldmath}

\begin{figure}[htb]
\centering
\includegraphics[width=80mm]{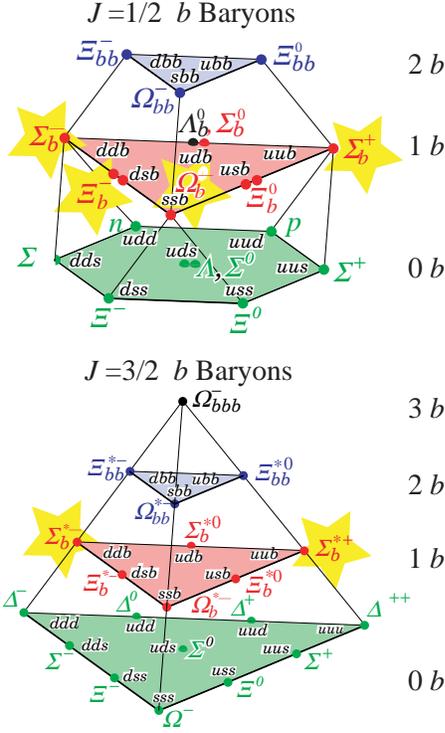}
\caption{
Multiplets of possible combinations of the heavy $b$ quark with
the $s$, $d$, and $u$ light quarks (figure adapted from Ref.~\cite{PDG08}).
}
\label{fig:bbaryon_mult}
\end{figure}

Until recently, the only directly observed 
$b$ baryon was the $\Lambda_b$~\cite{PDG08} 
(there is indirect evidence for $\Xi_b$ through
$\Xi$-lepton correlations at LEP~\cite{LEP_xib}).
As the Tevatron collects larger and larger data-sets,
the prospects improve for observing the rarer $b$ baryons
in decay channels providing good triggers.
If we consider combinations of only the heavier $b$ quark and the
three light quarks $s$, $d$, and $u$ quarks, multiplets of 
possible states are shown in Fig.~\ref{fig:bbaryon_mult}, very
similar to the multiplets for charm baryons.
Double-heavy baryons containing both a charm and $b$ quark are
not shown here.

We can also consider a heavy $b$ baryon as a
$L=0$ ``atomic" system if we take the heavy $b$ quark approximately at
rest in the rest frame of the $b$ baryon, and orbited by the 
{\em diquark} of the
two light quarks.
The two light quarks in the diquark can be anti-aligned, giving
diquark spin $s_{qq} = 0$ as for the ground state
$\Lambda_b^0$ ($bud$).
If the spins of the $u$ and $d$ quarks are instead aligned so
that $s_{qq} = 1$, it results in the more massive  $\Sigma^0_b$ baryon.
For the $L=0$ baryons, the total angular momentum is
$J = s_Q + s_{qq}$, where $s_Q$ is the spin of the heavy $b$ quark.
The two possibilities are therefore
the $J^P = \frac{1}{2}^+$ $\Sigma^0_b$ or the 
    $J^P = \frac{3}{2}^+$ $\Sigma^{*0}_b$ as shown in
Fig.~\ref{fig:bbaryon_mult}.

Such baryonic systems are valuable for testing predictions from
HQET, lattice gauge QCD, potential models, and sum rules.
The mass difference $M(\Sigma_b) - M(\Lambda_b)$ probes the 
energy from the diquark spin alignment,
$M(\Sigma^*_b) - M(\Sigma_b)$ the equivalent of hyperfine splitting,
and 
$M(\Sigma_b^-)[bdd] - M(\Sigma_b^+)[buu]$ the difference due
to isospin or effective mass of the $u$ and $d$ quarks.
Equivalent measures are also now accessible now that we have
evidence for and measurements of the properties of 
$b$ baryons containing strange quarks such as the 
$\Xi_b \thinspace (bsd)$ and the $\Omega_b \thinspace (bss)$.

\subsection{\boldmath Masses of $\Sigma^{(*)}_b$ Baryons
\unboldmath}

In 1.1~fb$^{-1}$ of data, the CDF Collaboration has
observed and measured the properties of all four
$\Sigma^{(*)\pm}_b$ charged baryons~\cite{CDF_Sigmab}.
The neutral states $\Sigma^{(*)0}_b$ would decay
into $\Lambda^0_b \pi^0$ which is difficult to reconstruct
at the Tevatron.
Starting with an optimized selection, they form a large,
clean sample of $\Lambda^0_b \rightarrow \Lambda_c^+ \pi^-$
decays where $\Lambda_c \rightarrow pK \pi$.
Since the $\Lambda^0_b$ decays weakly, decay length cuts
can be used to further reduce combinatorial background.
Since the decay of interest is a strong decay,  each of the
fitted  $N(\Lambda_b) = 3180 \pm 60$ candidates
is combined with a pion that is  constrained to originate from the 
primary vertex.
The largest background is combinatorial with random hadronization 
tracks being combined with $\Lambda^0_b$ baryons.

\begin{figure}[htb]
\centering
\includegraphics[width=80mm]{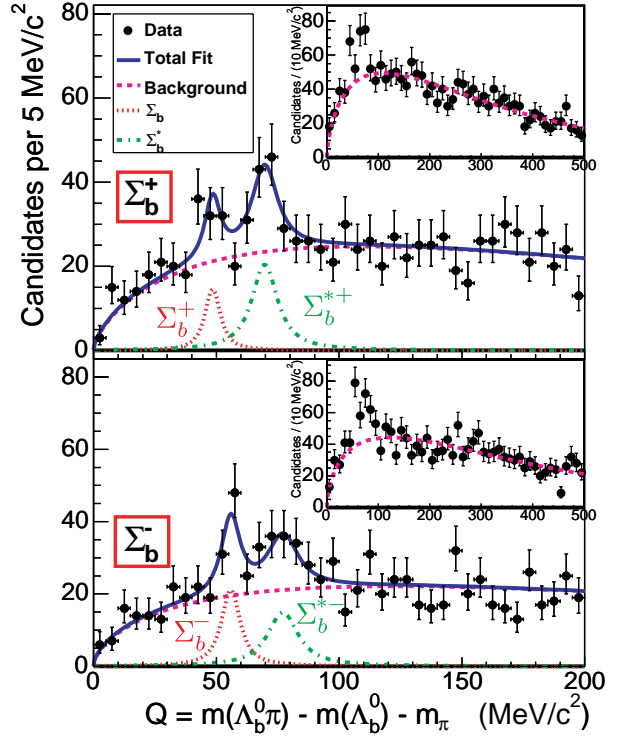}
\caption{$\Sigma^{(*)\pm}_b$:
$Q$ value distributions and fits for CDF analysis: top plot for
the $\Lambda^0_b \pi^+$ subsample that
contains $\Sigma^+_b$ and $\Sigma^{*+}_b$ and
bottom plot for
the $\Lambda^0_b \pi^-$ subsample that
contains $\Sigma^-_b$ and $\Sigma^{*-}_b$.
}
\label{fig:CDF_sigma}
\end{figure}

Figure~\ref{fig:CDF_sigma} shows the resulting $Q$ value mass
difference distributions for the 
subsequent samples.  The different charge sign
samples are kept separate and a simultaneous fit 
performed applying the
constraint that the mass splitting between the $J^P = \frac{1}{2}^+$
and $\frac{3}{2}^+$ states is the same:
$M(\Sigma{*^+}) - M(\Sigma^+_b) =
 M(\Sigma{*^-}) - M(\Sigma^-_b)$.
Both charge signs $\Sigma^-_b \thinspace (bdd)$ and
$\Sigma^+_b \thinspace (buu)$ show clear double peak structures,
with a 5.2$\sigma$ significance above background for signal.
Using the precision measurement of the 
$\Lambda^0_b$ mass~\cite{CDF_Lambdab}, the absolute masses 
are measured to be:
\begin{eqnarray}
M(\Sigma^+_b) & = & 5807.8 ^{+2.0}_{-2.2} \pm 1.7 
\thinspace {\mathrm{MeV}},
\nonumber \\
M(\Sigma^-_b) & = & 5815.2 \pm 1.0 \pm 1.7 
\thinspace {\mathrm{MeV}},
\nonumber \\
M(\Sigma^{*+}_b) & = & 5829.0 ^{+1.6 +1.7}_{-1.8 -1.8} 
\thinspace {\mathrm{MeV}},
\nonumber \\
M(\Sigma^{*-}_b) & = & 5836.4 \pm 2.0 ^{+1.8}_{-1.7}  
\thinspace {\mathrm{MeV}}.
\end{eqnarray}

These values can the be used to calculate various mass splittings
of interest as shown in Table~\ref{tab:mass_split} and compared
to expectations~\cite{CDF_Sigmab}.
Theoretical predictions are discussed thoroughly elsewhere
in these proceedings~\cite{karliner2} and show good agreement with
measurements.

\begin{table}[htb]
\begin{center}
\caption{
Comparison of measured CDF mass differences to expected 
values~\cite{CDF_Sigmab}.
}
\begin{tabular}{c|c|c}
\hline 
Property & \multicolumn{2}{c}{Values (MeV/$c^2$)}  \\
\cline{2-3}
         & Expected & Measured (CDF) \\
	 \hline \hline
Diquark spin alignment &  &  \\
$ M(\Sigma_b^+) - M(\Lambda^0_b)$ & 180 -- 210  & $
188.1^{+2.0 + 0.2 \dag}_{-2.2 - 0.3}$\\
$M(\Sigma_b^-) - M(\Lambda^0_b)$ & 
180 -- 210 & $195.5 \pm 1.0 \pm 0.2^{\dag}$ \\
(isospin averaged) &  & \\
$M(\Sigma_b) - M(\Lambda^0_b)$ & 194~\cite{karliner1} & 192 \\ \hline
Hyperfine splitting 
 & 10 -- 40 & $21.1^{+2.0+0.4}_{-1.9 - 0.3}$ \\
$M(\Sigma^*_b) - M(\Sigma_b)$  & $20.0 \pm 0.3$~\cite{karliner2} & \\ \hline
Isospin ($u,d$) diff. & & \\
$M(\Sigma^-_b) - M(\Sigma^+_b)$ & 5 -- 7 & $7.4 ^{+2.2^\dag}_{-2.4}$ \\ \hline
Widths & & \\
 $\Gamma(\Sigma_b), \Gamma(\Sigma^*_b)$ & $\sim 8$, $\sim 15$ & $-$\\
 \hline
 \multicolumn{3}{l}{$^{\dag}$ \footnotesize Calculated by author.}
 \end{tabular}
 \label{tab:mass_split}
 \end{center}
 \end{table}

\subsection{\boldmath Mass of $\Xi_b$ Baryon
\unboldmath}

The Tevatron has also recently discovered the first 
$b$ baryons containing strange quarks.  A possible
candidate $\Xi^0_b$ ($bsu$) should have dominant decays 
$\Xi^0_b \rightarrow \Xi^0_c \pi^0$ and $D^0 \Lambda$ with
neutral states difficult to reconstruct cleanly at the Tevatron.
A better candidate is the charged $b$ baryon
$\Xi^{\pm}_b$, first observed~\cite{D0_Xib} by the D\O\ Collaboration
in the decay $\Xi^{\pm}_b \rightarrow J/\psi \Xi^{\pm}$ and
and also observed~\cite{CDF_Xib} CDF Collaboration soon after
in the same decay mode plus the additional decay mode
$\Xi^{\pm}_b \rightarrow \Xi^0_c \pi^{\pm}$.
The $\Xi_b^{\pm}$ is expected to decay weakly with a lifetime
roughly comparable to other $b$ hadrons.

Both collaborations reconstruct the new state in the
decay chain $\Xi^{\pm}_b \rightarrow J/\psi \Xi^{\pm}$,
$J/\psi \rightarrow \mu^+ \mu^-$,
$\Xi \rightarrow \Lambda \pi$, $\Lambda \rightarrow p \pi$.
The reconstruction of the charged $\Xi^{\pm}$ with an average
decay length of approximately $5\thinspace {\mathrm{cm}}$ is challenging
in the tracking systems.  CDF uses silicon-only tracking, a first for
an experiment at a hadron collider, and also modified their
vertexing software to include this supplement. D\O\ reprocesses
tracks using special parameter settings to improve the efficiency
to reconstruct tracks with high impact parameters.

\begin{figure}[htb]
\centering
\includegraphics[width=80mm]{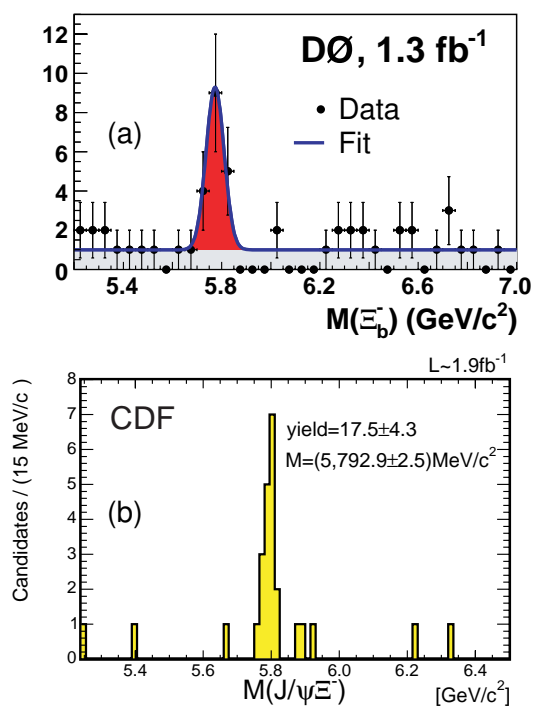}
\caption{
$\Xi^{\pm}_b$ signal mass peaks in $J/\psi \Xi^{\pm}$ invariant
mass distributions following selection cuts for
(a) the D\O\ Collaboration and (b) the CDF Collaboration.
}
\label{fig:Xib}
\end{figure}

In 1.1~fb$^{-1}$ of data, D\O\ reconstructs $1151 \pm 46$ $\Xi^{\pm}$
baryons, and in 1.9~fb$^{-1}$ of data, CDF reconstructs
$23500 \pm 340$ $\Xi^{\pm}$ baryons.
Further selection cuts based on momenta, decay lengths, and 
vertex quality with the 
$J/\psi$ are made, with selections optimized
on wrong-sign data and signal Monte Carlo samples by D\O, and
on a $B^+ \rightarrow J/\psi K^+$ control sample by CDF.
The resulting invariant mass distributions, including mass constraints
to the $J/\psi$ mass, are shown in Fig.~\ref{fig:Xib} with
signal yields of $15.2 \pm 4.4$ events (significance
of 5.5$\sigma$ above background) and
$17.5 \pm 4.3$ events (significance of 7.7$\sigma$ above
background) for D\O\ and CDF, respectively.
Fits to these distributions yield:
\begin{eqnarray}
M(\Xi_b) & = & 5774 \pm 11 \pm 15 \thinspace {\mathrm{MeV}}
\thinspace \thinspace {\mathrm{D\O}}, \\ \nonumber
M(\Xi_b) & = & 5792.9 \pm 2.4 \pm 1.7 \thinspace {\mathrm{MeV}}
\thinspace \thinspace {\mathrm{CDF}}.
\end{eqnarray}
D\O\ also observes a lifetime consistent with expectations.

\begin{figure}[htb]
\centering
\includegraphics[width=80mm]{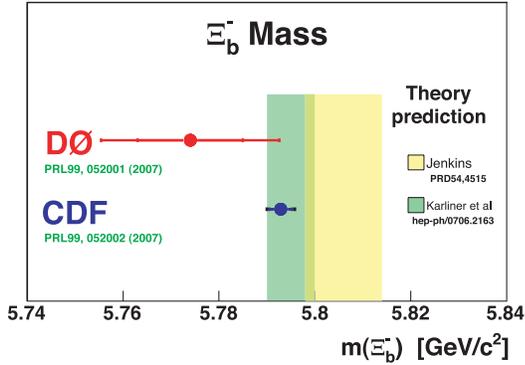}
\caption{
Measurements of the $\Xi^{\pm}_b$ mass compared to
theoretical predictions.
}
\label{fig:Xib_theory}
\end{figure}

These measurements show good agreement 
compared to theoretical predictions~\cite{Xib_theory} as shown
in Fig.~\ref{fig:Xib_theory}.

\subsection{\boldmath Mass of $\Omega_b$ Baryon
\unboldmath}

(Added in press.) Although not reported at the conference,
the D\O\ Collaboration has subsequently observed a
signal of $17.8 \pm 5.0$ events (statistical significance of
5.4$\sigma$ above background) and 
measured the mass of the doubly strange
$\Omega^-_b$ ($bss$) baryon~\cite{D0_Omegab} in 1.3~fb$^{-1}$ of
data.  Without details, the measured mass is:
\begin{equation}
M(\Omega^{\pm}_b) = 6165 \pm 10 \pm 13 \thinspace {\mathrm{MeV}}.
\end{equation}

\section{Conclusions and Prospects}

The renaissance of $b$ hadron spectroscopy (and measurement of
properties) continues as new $b$ baryon states continue to be
discovered and a number of excited $B$ meson states 
separated in a model-independent
way for the first time.  
The agreement of data to theory predictions is good for
most $B$ mesons, but less so for the 
orbitally excited $B^{**}$ and $B^{**}_s$ states. There is
excellent data-theory agreement for the new heavy $b$ baryons.
These measurements are providing useful input and comparisons to
potential models, HQET, lattice gauge calculations, and other
QCD models.  There are outstanding prospects for continued
{\em precision} predictions (e.g., $\Sigma^{(*)}_b$ inputs
for $\Xi_b$ predictions).

The performance of the Tevatron collider has been excellent, with
expectations to at least double the data-set to 
approximately 6--8~fb$^{-1}$ by the end of running in 
2009--2010.  There are very good prospects for increasing the sample
sizes of heavier $b$ hadron states and for the discovery of even
more states such as the double-heavy baryons.

\begin{acknowledgments}
I would like to acknowledge useful discussions with 
M.~Karliner.  I would also like to thank our hosts for their 
hospitality - the 11-course feast high up the world's tallest
building was particularly memorable, and I am now a life-long fan
of the work of Taiwanese sculpture Ju Ming.

\end{acknowledgments}

\bigskip % extra skip inserted
%\begin{thebibliography}{9}   % Use for  1-9  references

\end{document}